\documentclass[11pt,a4paper]{article}
\usepackage{amssymb}
\usepackage{graphicx}
\usepackage{amsmath}
\usepackage{makeidx}
\usepackage{indentfirst}
\usepackage[T1]{fontenc}
\usepackage[utf8]{inputenc}
\usepackage{authblk}

\newcounter{resultnum}[section]
\newcounter{conclusionnum}[section]
\newcounter{conditionnum}[section]
\newcounter{conjecturenum}[section]
\newcounter{examplenum}[section]
\newcounter{exercisenum}[section]
\newcounter{lemmanum}[section]
\newcounter{notationnum}[section]
\newcounter{theoremnum}[section]
\newcounter{definitionnum}[section]
\newcounter{corollarynum}[section]
\newcounter{remarknum}[section]
\newcounter{propositionnum}[section]
\newcounter{acknowledgementnum}[section]
\newcounter{algorithmnum}[section]
\newcounter{axiomnum}[section]
\newcounter{casenum}[section]
\newcounter{claimnum}[section]
\newcounter{summarynum}[section]
\newcounter{problemnum}[section]
\begin{document}

\title{Ghost--Free Massive $f(R)$ Theories Modelled as \\
Effective Einstein Spaces \& Cosmic Acceleration}
\date{October 15, 2014}

\author{{\Large \textbf{Sergiu I. Vacaru}}\thanks{%
sergiu.vacaru@uaic.ro} \\
{\qquad } \\
Theory Division, CERN, CH-1211, Geneva 23, Switzerland \thanks{%
associated visiting researcher}\\
and \\
Rector's Office, Alexandru Ioan Cuza University, \\
Alexandru Lapu\c sneanu street, nr. 14, UAIC -- Corpus R, office 323;\\
Ia\c si,\ Romania, 700057 }

\maketitle

\begin{abstract}
We study how massive ghost--free gravity $f(R)$--modified theories, MGFTs,
can be encoded into generic off--diagonal Einstein spaces. Using "auxiliary"
connections completely defined by the metric fields and adapted to
nonholonomic frames with associated  nonlinear connection structure, we
decouple and integrate in certain general forms the field equations in MGFT.
Imposing additional nonholonomic constraints, we can generate Levi--Civita,
LC, configurations and mimic MGFT effects via off--diagonal interactions of
effective Einstein and/or Einstein--Cartan gravity with nonholonomically
induced torsion. We show that imposing nonholonomic constraints it is
possible reproduce very specific models of massive $f(R)$ gravity studied in
\cite{sarid,odintsr}. The cosmological evolution of ghost--free
off--diagonal Einstein spaces is investigated. Certain compatibility of MGFT
cosmology to small off--diagonal deformations of $\Lambda $CDM models is
established.
\end{abstract}

\affil[1]{\small Theory Division, CERN, CH-1211, Geneva 23, Switzerland \footnote{associated visiting researcher};\ and
\newline
 Rector's Office, Alexandru Ioan Cuza University,\
 Alexandru Lapu\c sneanu street, \newline  nr. 14, UAIC - Corpus R, office 323;\
 Ia\c si,\ Romania, 700057
}

\textit{Introduction:}\textbf{\ }In \cite{sarid,odintsr}, two models of
nonlinear massive gravitational theories including $f(R)$ modifications were
elaborated. Such theories contain the benefits of the dRGT model \cite{drgt}
and are free of ghost modes \cite{ghost}. Advantages are  that
by tuning the $f(R)$ functional (on such modifications, see reviews \cite%
{mgt}), we can stabilize cosmological backgrounds, elaborate various types
cosmological evolution scenarios, unified description of inflation and
late--time acceleration etc. The main goal of \cite{sarid} is to perform a
general analysis for arbitrary $f(R)$ theory but the references \cite%
{odintsr} provide solutions for explicit cosmological problems of such
theories. From general theoretical considerations, the $f(R)$ paradigm  attempts to explain the universe acceleration and
dark energy/ matter problems through infra--red (IR) modifications of the
general relativity (GR) theory and understanding possible physical
implications of the massive spin--2 theory. In this paper, we generate a
very specific model of massive $f(R)$ gravity constraining nonholonomically
the corresponding system of modified gravitational equations. We shall
analyze possible cosmological implications for such special cases containing
small off--diagonal corrections. On the other hand, ultra-violet (UV)
corrections expected to be of quantum origin (see Ref. \cite{stelle} on
possible effective actions). Cosmological implications of massive gravity
were also analyzed in the framework of modified gravity theories, MGT, \cite%
{mgt1},  also cosmological models related to bi--metric gravity \cite%
{bmcm}.

It is the point of this paper to apply in MGFT the so-called anholonomic
frame deformation method, AFDM, \cite{afdm} for constructing generic
off--diagonal exact solutions. Such a method provides a geometric techniques
which allows us to integrate systems of partial differential equations,
PDEs, with functional and parametric dependencies for the Levi--Civita (zero
torsion) and nontrivial torsion configurations.

\vskip5pt

\textit{The geometric setup:} We shall work on a pseudo-Riemannian manifold $%
V,$\ $\dim V=4,$  where a Whitney sum $\mathbf{N}$ is defined
for its tangent space $TV,$ $\mathbf{N}:\ TV=hTV\oplus vTV.$ Such a
decomposition defines a nonholonomic (equivalently, non-integrable, or
anholonomic) horizontal (h) and vertical (v) splitting, i.e. a nonlinear
connection (\textit{N-connection}) structure, see details in \cite{afdm}.
The local coefficients $\{N_{i}^{a}(u)\},$ where $\mathbf{N}%
=N_{i}^{a}(x,y)dx^{i}\otimes \partial /\partial y^{a}$ for certain local
coordinates $u=(x,y),$ or $u^{\alpha }=(x^{i},y^{a}),$ with $h$-indices $%
i,j,=1,2$ and $v$-indices $a,b,...=3,4,$\footnote{%
We shall use the Einstein rule on summation on \textquotedblright
up-low\textquotedblright\ cross indices. Boldface letters are written in
order to emphasize that a N-connection spitting is considered on a manifold $%
\mathbf{V=(}V,\mathbf{N).}$} define naturally N-adapted frame and,
respectively, dual frame structures, $\mathbf{e}_{\nu }=(\mathbf{e}%
_{i},e_{a})$ and $\mathbf{e}^{\mu }=(e^{i},\mathbf{e}^{a}),$ where
\begin{equation}
\mathbf{e}_{i}=\partial /\partial x^{i}-\ N_{i}^{a}(u)\partial /\partial
y^{a},\ e_{a}=\partial _{a}=\partial /\partial y^{a},\mbox{ and  }%
e^{i}=dx^{i},\ \mathbf{e}^{a}=dy^{a}+\ N_{i}^{a}(u)dx^{i}.  \label{nadif}
\end{equation}%
In general, such local (co) bases are nonholonomic, i.e. $[\mathbf{e}%
_{\alpha },\mathbf{e}_{\beta }]=\mathbf{e}_{\alpha }\mathbf{e}_{\beta }-%
\mathbf{e}_{\beta }\mathbf{e}_{\alpha }=W_{\alpha \beta }^{\gamma }\mathbf{e}%
_{\gamma }$ with anholonomy coefficients $W_{ia}^{b}=\partial
_{a}N_{i}^{b},W_{ji}^{a}=\Omega _{ij}^{a}=\mathbf{e}_{j}\left(
N_{i}^{a}\right) -\mathbf{e}_{i}(N_{j}^{a}),$ where $\Omega _{ij}^{a}$ is
the N-connection curvature. With respect to (\ref{nadif}), any metric tensor
$\mathbf{g}$ can be expressed as a distinguished metric, d--metric,
\begin{equation}
\mathbf{g}=g_{\alpha }(u)\mathbf{e}^{\alpha }\otimes \mathbf{e}^{\beta
}=g_{i}(x^{k})dx^{i}\otimes dx^{i}+g_{a}(x^{k},y^{b})\mathbf{e}^{a}\otimes
\mathbf{e}^{a}.  \label{dm1}
\end{equation}

For any prescribed N--connection and d--metric structures, we can work
equivalently with two linear connections,
\begin{equation*}
(\mathbf{g,N})\rightarrow \left\{
\begin{array}{cc}
\mathbf{\nabla :} & \mathbf{\nabla g}=0;\ ^{\nabla }\mathcal{T}=0; \\
\widehat{\mathbf{D}}: & \widehat{\mathbf{D}}\mathbf{g}=0;\ h\widehat{%
\mathcal{T}}=0,v\widehat{\mathcal{T}}=0,hv\widehat{\mathcal{T}}\neq 0,%
\end{array}%
\right.
\end{equation*}%
where $\mathbf{\nabla }$ is the torsionless Levi--Civita, LC, connection and
$\widehat{\mathbf{D}}=h\widehat{\mathbf{D}}+v\widehat{\mathbf{D}}$ is the
so--called canonical distinguished connection, d--connection. The value $%
\widehat{\mathbf{D}}$ preserves the $h$-$v$--splitting under parallel
transports but $\mathbf{\nabla }$ does not have such a property.
Nevertheless, there is a canonical distortion distinguished tensor,
d-tensor, $\widehat{\mathbf{Z}}=\{\widehat{\mathbf{T}}_{\ \beta \gamma
}^{\alpha }\},$ which is an algebraic combination of the coefficients of the
corresponding torsion d-tensor $\widehat{\mathcal{T}}=\{\widehat{\mathbf{T}}%
_{\ \beta \gamma }^{\alpha }\}.$ This defines a canonical distortion
relation $\widehat{\mathbf{D}}=\nabla +\widehat{\mathbf{Z}}$ which is
adapted to the N--splitting. The torsions, $\widehat{\mathcal{T}}$ and $\
^{\nabla }\mathcal{T}=0,$ and curvatures, $\widehat{\mathcal{R}}=\{\widehat{%
\mathbf{R}}_{\ \beta \gamma \delta }^{\alpha }\}$ and $\ ^{\nabla }\mathcal{R%
}=\{R_{\ \beta \gamma \delta }^{\alpha }\},$ respectively, of $\widehat{%
\mathbf{D}}$ and $\nabla $ can be defined and computed in standard
coordinate free and/or coefficient forms.

The Ricci tensors of $\widehat{\mathbf{D}}$ and $\nabla $ are defined $\
\widehat{\mathcal{R}}ic=\{\widehat{\mathbf{R}}_{\ \beta \gamma }:=\widehat{%
\mathbf{R}}_{\ \alpha \beta \gamma }^{\gamma }\}$ and $Ric=\{R_{\ \beta
\gamma }:=R_{\ \alpha \beta \gamma }^{\gamma }\}.$ For instance, the Ricci
d-tensor $\widehat{\mathcal{R}}ic$ is characterized by four subsets of $h$-$%
v $ N-adapted coefficients,
\begin{equation}
\widehat{\mathbf{R}}_{\alpha \beta } = \{\widehat{R}_{ij}:=\widehat{R}_{\
ijk}^{k},\ \widehat{R}_{ia}:=-\widehat{R}_{\ ika}^{k},\ \widehat{R}_{ai}:=%
\widehat{R}_{\ aib}^{b},\ \widehat{R}_{ab}:=\widehat{R}_{\ abc}^{c}\}.
\label{driccic}
\end{equation}

Alternatively to the LC-scalar curvature, $\ R:=\mathbf{g}^{\alpha \beta
}R_{\alpha \beta },$ we can introduce the scalar of canonical d--curvature, $%
\ \widehat{\mathbf{R}}:=\mathbf{g}^{\alpha \beta }\widehat{\mathbf{R}}%
_{\alpha \beta }=g^{ij}\widehat{R}_{ij}+g^{ab}\widehat{R}_{ab}.$\footnote{%
Any (pseudo) Riemannian geometry can be equivalently described by both
geometric data $\left( \mathbf{g,\nabla }\right) $ and $(\mathbf{g,N,}%
\widehat{\mathbf{D}})$, where the canonical distortion relations $\widehat{%
\mathcal{R}}=\ ^{\nabla }\mathcal{R+}\ ^{\nabla }\mathcal{Z}$ and $\widehat{%
\mathcal{R}}ic=Ric+\widehat{\mathcal{Z}}ic,$ with respective distortion
d-tensors $\ ^{\nabla }\mathcal{Z}$ and $\widehat{\mathcal{Z}}ic,$ are
computed for $\widehat{\mathbf{D}}=\nabla +\widehat{\mathbf{Z}}.$ To prove
the decoupling of fundamental gravitational equations in general relativity,
GR, and various MGFTs is possible for d-metrics and the canonical
d--connection working with respect to N-adapted frames. LC-configurations
can be extracted from certain classes of solutions of (modified)
gravitational field equations if additional conditions are imposed,
resulting in zero values for the canonical d-torsion, $\widehat{\mathcal{T}}%
=0$.}

\vskip5pt

\textit{Field equations in MGFT and N--adapted variables:}\ We follow the
model elaborated in \cite{odintsr} and reformulate it on a nonholonomic
manifold $\mathbf{V}$ enabled with N--connection structure $\mathbf{N}$ and
two d--metrics where $\mathbf{g=\{g}_{\alpha \beta }\mathbf{\}}$ is the
dynamical d--metric and $\mathbf{q}=\{\mathbf{q}_{\alpha \beta }\}$ is the
so--called non--dynamical reference metric. In our approach, we work with $%
\widehat{\mathbf{D}}$ instead of $\nabla$ and $\widehat{\mathbf{R}}$ is
computed for $\mathbf{g,}$ the nonzero graviton mass is denoted by $\mu ,$ $%
M_{P}$ is the Planck mass.\footnote{%
Our system of "N--adapted notations" is similar to that considered in \cite%
{vmgbmconf,elizaldev}.}

Let us consider the d--tensor $(\sqrt{\mathbf{g}^{-1}\mathbf{q}})_{~\nu
}^{\mu }$ computed as the square root of $\mathbf{g}^{\mu \rho }\mathbf{q}%
_{\rho \nu },$ where
\begin{equation*}
(\sqrt{\mathbf{g}^{-1}\mathbf{q}})_{~\rho }^{\mu }(\sqrt{\mathbf{g}^{-1}%
\mathbf{q}})_{~\nu }^{\rho }=\mathbf{g}^{\mu \rho }\mathbf{q}_{\rho \nu }, %
\mbox{ and } \ \sum\limits_{k=0}^{4}~^{k}\beta ~e_{k}(\sqrt{\mathbf{g}^{-1}%
\mathbf{q}})=3-tr\sqrt{\mathbf{g}^{-1}\mathbf{q}}-\det \sqrt{\mathbf{g}^{-1}%
\mathbf{q}},
\end{equation*}
for some coefficients $~^{k}\beta .$ The values $e_{k}(\mathbf{X})$ can be
defined for any d--tensor $\mathbf{X}_{~\rho }^{\mu }$ and trace $X=[X]:=tr(%
\mathbf{X})=\mathbf{X}_{~\mu }^{\mu },$ where
\begin{eqnarray*}
e_{0}(X) &=&1,e_{1}(X)=X,2e_{2}(X)=X^{2}-[X^{2}],\ 6e_{3}(X) =
X^{3}-3X[X^{2}]+2[X^{3}], \\
24e_{4}(X) &=&X^{4}-6X^{2}[X^{2}]+3[X^{2}]^{2}+8X[X^{3}]-6[X^{4}];\ e_{k}(X)
= 0 \mbox{ for }k>4.
\end{eqnarray*}%
We shall use also the mass--deformed scalar curvature  $\mathbf{\tilde{R}}%
:=\ \widehat{\mathbf{R}}+2~\mu ^{2}(3-tr\sqrt{\mathbf{g}^{-1}\mathbf{q}}%
-\det \sqrt{\mathbf{g}^{-1}\mathbf{q}})$.

The action $\mathcal{S}$ for MGFT is postulated in the form
\begin{equation}
\mathcal{S}=M_{P}^{2}\int d^{4}u\sqrt{|\mathbf{g}|}[\mathbf{f}(\mathbf{%
\tilde{R}})+~^{m}\mathcal{L}],  \label{actmgft}
\end{equation}%
where $~^{m}\mathcal{L}(\mathbf{g,N})$ is the Lagrange density for the
matter fields.\footnote{%
For simplicity, we consider matter actions $~^{m}\mathcal{S}=\int d^{4}u%
\sqrt{|\mathbf{g}|}~^{m}\mathcal{L}$ which only depend on the coefficients
of a metric field and not on their derivatives. Here we note that the
geometric constructions in this paper can also be performed in similar form
for cosmological models \cite{odintsr} but must be supplemented with a number of formulas would
contain nonholonomic constraints for additional physical assumptions. To
work with the action (\ref{actmgft}) is a more convenient choice for
emphasizing in an "economic" way all priorities of our geometric approach.}
The energy--momentum d--tensor can be computed via N--adapted variational
calculus, {\small
\begin{equation}
\ ^{m}\mathbf{T}_{\alpha \beta }:=-\frac{2}{\sqrt{|\mathbf{g}_{\mu \nu }|}}%
\frac{\delta (\sqrt{|\mathbf{g}_{\mu \nu }|}\ \ ^{m}\mathcal{L})}{\delta
\mathbf{g}^{\alpha \beta }} = \ ^{m}\mathcal{L}\mathbf{g}^{\alpha \beta }+2%
\frac{\delta (\ ^{m}\mathcal{L})}{\delta \mathbf{g}_{\alpha \beta }}.
\label{emdt}
\end{equation}%
} Applying such a calculus to $\mathcal{S}$ (\ref{actmgft}), with $\ ^{1}%
\mathbf{f}(\mathbf{\tilde{R}}):=d\mathbf{f}(\mathbf{\tilde{R}})/d\mathbf{%
\tilde{R},}$ see details in \cite{afdm}, we obtain the field equations
\begin{equation}
\widehat{\mathbf{R}}_{\mu \nu }=\mathbf{\Upsilon }_{\mu \nu },  \label{mfeq}
\end{equation}%
where $\mathbf{\Upsilon }_{\mu \nu }=~^{m}\mathbf{\Upsilon }_{\mu \nu }+~^{f}%
\mathbf{\Upsilon }_{\mu \nu }+~^{\mu }\mathbf{\Upsilon }_{\mu \nu },$ for%
\begin{eqnarray}
~^{m}\mathbf{\Upsilon }_{\mu \nu } & =&\frac{1}{2M_{P}^{2}}\ ^{m}\mathbf{T}%
_{\alpha \beta },\ \ ^{f}\mathbf{\Upsilon }_{\mu \nu }= (\frac{\mathbf{f}}{%
2~^{1}\mathbf{f}}-\frac{\widehat{\mathbf{D}}^{2}\ ^{1}\mathbf{f}}{~^{1}%
\mathbf{f}})\mathbf{g}_{\mu \nu }+\frac{\widehat{\mathbf{D}}_{\mu }\widehat{%
\mathbf{D}}_{\nu }\ ^{1}\mathbf{f}}{~^{1}\mathbf{f}},  \label{source} \\
~^{\mu }\mathbf{\Upsilon }_{\mu \nu } &=& -2\mu ^{2}[(3-tr\sqrt{\mathbf{g}%
^{-1}\mathbf{q}}-\det \sqrt{\mathbf{g}^{-1}\mathbf{q}}) - \frac{1}{2}\det
\sqrt{\mathbf{g}^{-1}\mathbf{q}})]\mathbf{g}_{\mu \nu }  \notag \\
&&+ \frac{\mu ^{2}}{2}\{\mathbf{q}_{\mu \rho }[(\sqrt{\mathbf{g}^{-1}\mathbf{%
q}})^{-1}]_{~\nu }^{\rho }+\mathbf{q}_{\nu \rho }[(\sqrt{\mathbf{g}^{-1}%
\mathbf{q}})^{-1}]_{~\mu }^{\rho }\}.  \notag
\end{eqnarray}

We note that the Bianchi identities for the data $(\mathbf{g,N,}\widehat{%
\mathbf{D}})$ are given by introducing nonholonomic deformations $\nabla =$ $%
\widehat{\mathbf{D}}-\widehat{\mathbf{Z}}$ into the standard relations $%
\nabla ^{\alpha }(R_{\alpha \beta }-\frac{1}{2}g_{\alpha \beta }R)=0$ and $%
\nabla ^{\alpha }T_{\alpha \beta }=0.$ Even, in general, $\widehat{\mathbf{D}%
}^{\alpha }\mathbf{T}_{\alpha \beta }=\mathbf{Q}_{\beta }\neq 0,$ such a $%
\mathbf{Q}_{\beta }[\mathbf{g,N}]$ is completely defined by the d--metric
and chosen N--connection structure. This is a consequence of the
nonholonomic structure. A similar "problem" exists in Lagrange mechanics
with non--integrable constraints when the standard conservation laws do not
hold true. A new class of effective variables can be introduced using
Lagrange multiples. We omit cumbersome formulas for the Bianchi densities
and conservation laws with nonholonomic constraints written in variables $(%
\mathbf{g,N,}\widehat{\mathbf{D}}).$

\vskip5pt

\textit{Encoding and decoupling properties of field equations in MGFT:}\ The
generalized gravitational field equations written with respect to N--adapted
frames (\ref{mfeq}) are similar to those studied in our works \cite%
{afdm,vmgbmconf,elizaldev}. The main difference of such MTGs is determined
by a corresponding source which in this work is considered in the form (\ref%
{source}). Applying the AFDM, we can construct very general classes of
generic off--diagonal solutions which encode both $f$--modifications and
massive gravity effects with nonzero $\mu .$

For simplicity, we shall consider nonholonomic dynamical systems in MGTs
which via frame transforms and connection deformations can be transformed
into certain effective off--diagonal Einstein manifolds described by
d--metrics with one Killing symmetry on $\partial /\partial y^{3},$ i.e. the
gravitational and matter fields do not depend on variable $y^{4}.$\footnote{%
It is possible to construct metrics with non--Killing symmetries depending
on all spacetime coordinates. This requests a more advanced and cumbersome
geometric techniques, see examples in \cite{afdm,elizaldev} and references
therein.} This is described by ansatz (\ref{dm1}) with
\begin{equation}
g_{i}=e^{\psi (x^{i})},g_{a}=h_{a}(x^{i},t),\
N_{i}^{3}=n_{i}(x^{k}),N_{i}^{4}=w_{i}(x^{k},t).  \label{param1}
\end{equation}
The effective source is chosen for a timelike coordinate $y^{4}=t,$ where
\begin{eqnarray}
\mathbf{\Upsilon }_{\mu \nu } &\rightarrow &\mathbf{\hat{\Upsilon}}_{\mu \nu
}=diag[\Upsilon _{1}=\Upsilon _{2}, \Upsilon _{2}= ~^{m}\tilde{\Upsilon}%
(x^{i})+~^{f}\tilde{\Upsilon}(x^{i})+~^{\mu }\tilde{\Upsilon}(x^{i}),  \notag
\\
&& \Upsilon _{3} =\Upsilon _{4},\ \Upsilon _{4}=~^{m}\Upsilon (x^{i},t)+\
^{f}\Upsilon (x^{i},t)+\ ^{\mu }\Upsilon (x^{i},t)]  \notag \\
&\rightarrow &\mathbf{\check{\Upsilon}}_{\mu \nu }=(~^{m}\check{\Lambda}%
+~^{f}\check{\Lambda}+~^{\mu }\check{\Lambda}\ )\mathbf{g}_{\alpha \beta }.\
\label{dsours}
\end{eqnarray}%
The assumption for the first parametrization in (\ref{dsours}) is that the
matter fields and effective sources, $\mathbf{\Upsilon }_{\mu ^{\prime }\nu
^{\prime }}=e_{~\mu ^{\prime }}^{\mu }e_{~\nu ^{\prime }}^{\nu }\mathbf{\hat{%
\Upsilon}}_{\mu \nu },$ are generated in N--adapted frames by two types
functions/ distributions $\ \tilde{\Upsilon}(x^{i})$ and $\Upsilon
(x^{i},t). $ The left labels refer to contributions in such sources by $f$%
--modifications and/or by mass $\mu $--modifications. In general, we get 4
independent N--adapted coefficients of $\mathbf{\Upsilon }_{\mu \nu }=diag\{%
\mathbf{\Upsilon }_{\mu }(x^{i},t)\}$ for variations in (\ref{emdt}) using (%
\ref{param1}). For cosmological applications, we can model sources of matter
fields by an energy--momentum tensor for ideal fluids as in GR but with
generic off--diagonal metrics\footnote{%
such metrics can not be diagonalized by coordinate transforms because for
general N--connections the anholonomy coefficients, $W_{\alpha \beta
}^{\gamma },$ are not zero} (encoding contributions from MGT). In N--adapted
frames,
\begin{equation}
\mathbf{T}_{\alpha \beta }=p\mathbf{g}_{\alpha \beta }+(\rho +p)\mathbf{v}%
_{\alpha }\mathbf{v}_{\beta }  \label{dsourc}
\end{equation}%
is defined for certain (effective) energy, $\rho ,$ and pressure, densities,
$p,$ respectively, $\widehat{\mathbf{v}}_{\alpha }$ being the four-velocity
of the fluid for which $\mathbf{v}_{\alpha }\mathbf{v}^{\alpha }=-1$ and $%
\mathbf{v}^{\alpha }=(0,0,0,1)$ in N-adapted comoving frames/coordinates.

A tedious calculus of N--adapted coefficients of the Ricci d--tensor for $%
\widehat{\mathbf{D}}$ computed for ansatz (\ref{param1}) and source (\ref%
{dsours}) transform (\ref{mfeq}) into a system of nonlinear PDEs:
\begin{eqnarray}
\psi ^{\bullet \bullet }+\psi ^{\prime \prime } = 2~(~^{m}\tilde{\Upsilon}%
+~^{f}\tilde{\Upsilon}+~^{\mu }\tilde{\Upsilon})&=&2~\tilde{\Upsilon},
\label{eq1} \\
\phi ^{\diamond }h_{3}^{\diamond } = 2h_{3}h_{4}~(~^{m}\Upsilon +\
^{f}\Upsilon +\ ^{\mu }\Upsilon )&=&2h_{3}h_{4}~\Upsilon,  \label{eq2} \\
n_{i}^{\diamond \diamond }+\gamma n_{i}^{\diamond } =0,~\beta w_{i}-\alpha
_{i}&=&0,  \label{eq4}
\end{eqnarray}%
for $\ \alpha _{i}=h_{3}^{\diamond }\partial _{i}\phi ,\beta
=h_{3}^{\diamond }\ \phi ^{\diamond },\gamma =\left( \ln
|h_{3}|^{3/2}/|h_{4}|\right) ^{\diamond },$ where
\begin{equation}
{\phi =\ln |h_{3}^{\diamond }/\sqrt{|h_{3}h_{4}|}|,\mbox{ and/ or }}\Phi
:=e^{{\phi }},  \label{genf}
\end{equation}%
is considered as a generating function. In above formulas, we use notations for partial derivatives: $\psi ^{\bullet }=\partial _{1}\psi
=\partial \psi /\partial x^{1},\psi ^{\prime }=\partial _{2}\psi ,{%
h_{3}^{\diamond }=}\partial _{4}{h_{3}.}$ For simplicity, we do not study in
this paper d--metrics for which ${h_{a}^{\diamond }=0}$ and/or $\mathbf{%
\Upsilon }_{\mu }=0$ (such solutions in vacuum MGFT can be constructed, for
instance, for $f$- and/or $\mu $-modifications of black hole solutions, see
examples in \cite{afdm}). Here we note that the relevant equations (\ref{eq2}), (\ref{eq4}) and respective coefficients can be computed in a similar form if
corresponding coordinates and indices are changed as $3\to 5$ and $4\to 6$,
which allows to extend the method for extra dimensions. Such recurrent
formulas can be proven for arbitrary finite number of extra (non) holonomic
coordinates. For simplicity, we analyse in this work only examples of
off--diagonal metrics for 4-d spacetimes.

The torsionless (Levi--Civita, LC) conditions are satisfied if there are
additionally imposed the conditions
\begin{equation}
w_{i}^{\diamond } = (\partial _{i}-w_{i}\partial _{4})\ln \sqrt{|h_{4}|}%
,(\partial _{i}-w_{i}\partial _{4})\ln \sqrt{|h_{3}|}=0,\ \partial _{k}w_{i}
=\partial _{i}w_{k},n_{i}^{\diamond }=0,\partial _{i}n_{k}=\partial
_{k}n_{i}.  \label{lccondb}
\end{equation}

\vskip5pt

\textit{Exact off--diagonal solutions in MGFT:}\ The system (\ref{eq1})-(\ref%
{eq4}) posses an important property when 1) $\psi $ is the solution of two
dimensional (2-d) Poisson equation with source $2(...)(x^{k});$ $h_{3}$ and $%
h_{4}$ are related to $\phi $ and sources via equation (\ref{genf}) and the
N--connection coefficients are determined correspondingly by integrating two
times on $t$ the equations for $n_{i}$ and from a system of first order
algebraic equations for $w_{i}.$ For MGTs, the procedure of finding locally
anisotropic and inhomogeneous cosmological solutions is described in \cite%
{elizaldev}.

We  fix the sum of nontrivial constants $\check{\Lambda}=~^{m}\check{%
\Lambda}+~^{f}\check{\Lambda}+~^{\mu }\check{\Lambda}$ and re--define the
generating function, $\Phi \longleftrightarrow \check{\Phi},$ using formulas
{\small
\begin{equation}
\check{\Lambda} \check{\Phi}^{2}=\left[ \Phi ^{2}|\Upsilon |+\int dt\ \Phi
^{2}|\Upsilon |^{\diamond }\right] ,\ \Phi ^{2}=\frac{|\check{\Lambda}|}{%
|\Upsilon |^{2}}\int dt\ \check{\Phi}^{2}|\Upsilon |,  \label{genfunct}
\end{equation}%
} where $(\Phi ^{2})^{\diamond }/|\Upsilon |=(\check{\Phi}^{2})^{\diamond }/%
\check{\Lambda}.$ In order to solve the second equation in (\ref{lccondb}), $%
(\partial _{i}-w_{i}\partial _{4})\ln \sqrt{|h_{3}|}=0,$ the generating
function $\Phi $ must be chosen to satisfy the conditions $(\partial
_{i}\Phi )^{\diamond }=\partial _{i}\Phi ^{\diamond }.$ We can parameterize
the solutions for the system (\ref{eq2}) and (\ref{genf}) in the form $h_{3}[%
\check{\Phi}]=\frac{\check{\Phi}^{2}}{4|\check{\Lambda}|}$ and $\ h_{4}[%
\check{\Phi}]=\frac{(\check{\Phi}^{\diamond })^{2}}{\check{\Lambda}\Phi ^{2}}%
=\frac{|\check{\Phi}^{\diamond }~\Upsilon |^{2}}{\check{\Lambda}|\check{%
\Lambda}|\int dt\ \check{\Phi}^{2}\ |\Upsilon |}.$

We find in explicit form solutions of algebraic equations in (\ref{eq4}) and
the conditions $\partial _{k}w_{i}=\partial _{i}w_{k}$ from the second line
in (\ref{lccondb}) if
\begin{equation}
w_{i}=\partial _{i}\Phi /\Phi ^{^{\diamond }}=\partial _{i}\widetilde{A},
\label{wcoef}
\end{equation}%
with a nontrivial function $\widetilde{A}(x^{k},t)$ depending on generating
function $\Phi $ via a first order Pfaff system. Integrating two times on $t$
in (\ref{eq4}), we express
\begin{equation}
n_{k}=\ _{1}n_{k}+\ _{2}n_{k}\int dy^{4}\ h_{4}/(\sqrt{|h_{3}|})^{3},
\label{ncoef}
\end{equation}%
where $\ _{1}n_{k}(x^{i})$ and $\ _{2}n_{k}(x^{i})$ are integration
functions. To generate LC-configurations we take $\ _{2}n_{k}=0$ and $%
\ _{1}n_{k}=\partial _{k}n(x^{i}).$

Putting together above formulas, we conclude that generic off--diagonal
quadratic elements
\begin{eqnarray}
ds^{2} &=& e^{\psi (x^{k},[~^{m}\tilde{\Upsilon}+~^{f}\tilde{\Upsilon}%
+~^{\mu }\tilde{\Upsilon}])}[(dx^{1})^{2}+(dx^{2})^{2}] +\frac{\check{\Phi}%
^{2}[dy^{3}+\partial _{k}n~dx^{k}]^{2}}{4|~^{m}\check{\Lambda}+~^{f}\check{%
\Lambda}+~^{\mu }\check{\Lambda}|}\pm  \notag \\
&&\frac{|[\check{\Phi}^{\diamond }]~[~^{m}\Upsilon +\ ^{f}\Upsilon +\ ^{\mu
}\Upsilon ]|^{2}}{|~^{m}\check{\Lambda}+~^{f}\check{\Lambda}+~^{\mu }\check{%
\Lambda}|^{\frac{3}{2}}\int dt\ \check{\Phi}^{2}\ |~^{m}\Upsilon +\
^{f}\Upsilon +\ ^{\mu }\Upsilon |} (dt+\partial _{i}\widetilde{A}[\check{\Phi%
}]~dx^{i})^{2}.  \label{excosms}
\end{eqnarray}%
determine generic off--diagonal solutions of the field equations in MGFT.
For well--defined assumptions on Killing symmetry on $\partial _{3}$ and
imposed at the end zero torsion conditions such metrics belong to the
integral variety of the system (\ref{eq1})-(\ref{lccondb}). We can generate
exact solutions in "pure" $f$--modified gravity if put $\ ^{\mu }\Upsilon =\
^{\mu }\Lambda =0$. If $\Lambda \neq 0$, we can nonholonomically induce a
nontrivial $\ ^{\mu }\Upsilon $. Inverse nonlinear transforms are possible
if we change mutually the left labels $\mu$ with $f$.

It should be noted that above classes of metrics can be extended to describe
exact solutions with nonholonomically induced torsion $\widehat{\mathcal{T}}%
=\{\widehat{\mathbf{T}}_{\ \beta \gamma }^{\alpha }[\check{\Phi},\tilde{%
\Upsilon},\Upsilon ,\check{\Lambda}]\}$ of $\widehat{\mathbf{D}}.$ We  substitute in (\ref{excosms}) $\partial _{k}n\rightarrow n_{k}(x^{i},t)$ (%
\ref{ncoef}) and take instead of (\ref{wcoef}) the value $w_{i}=\partial
_{i}\Phi /\Phi ^{^{\diamond }}.$ It is possible to re-write all coefficients
in terms of the generating function $\Phi ,$ or in terms of $\check{\Phi}.$
The LC conditions (\ref{lccondb}) are not satisfied for such configurations.%
\footnote{%
Such torsion fields are different from those in Einstein--Cartan, gauge
and/or string gravity where additional field equations and sources are
considered to define the torsion dynamics.}

\vskip5pt

\textit{On properties of off--diagonal solutions in MGFT and GR: } The
metrics (\ref{excosms}) describe locally anisotropic and inhomogeneous
spacetimes determined by certain classes of generating functions $\check{\Phi%
}(x^{i},t)$ and \newline
$\psi (x^{k},[~^{m}\tilde{\Upsilon}+~^{f}\tilde{\Upsilon}+~^{\mu }\tilde{%
\Upsilon}])$; sources $~^{m}\Upsilon (x^{i},t),\ ^{f}\Upsilon (x^{i},t),\
^{\mu }\Upsilon (x^{i},t)$ and $~^{m}\tilde{\Upsilon}(x^{i}),$ $~^{f}\tilde{%
\Upsilon}(x^{i}),~^{\mu }\tilde{\Upsilon}(x^{i})$, and integration functions
like $\partial _{k}n(x^{k})$; and effective cosmological constants $~^{m}%
\check{\Lambda},~^{f}\check{\Lambda},~^{\mu }\check{\Lambda},$ which can be
considered as integration constants. These values and one of the  $\pm $
should be fixed  such  that they are compatible with observational
data. We can generate inhomogeneous cosmological metrics taking certain
limits $\check{\Phi}(x^{i},t)\rightarrow \check{\Phi}(t)$ and for respective
sources ${\Upsilon}(x^{i},t)\rightarrow {\Upsilon}(t)$. Such solutions
generalize the class of known anisotropic solutions of Bianchi cosmology to
configurations;  the coefficients of metrics are not subject to typical
symmetric conditions for those spacetimes and, in our approach, may encode
geometric and physical data for MGFT interactions.

Fixing, for instance, $~^{\mu }\check{\Lambda}=~^{\mu }\tilde{\Upsilon}=\
^{\mu }\Upsilon =0,$ i.e. for the zero mass of graviton, the metrics (\ref%
{excosms}) reproduce certain results of $f(\ \widehat{\mathbf{R}})$ gravity
and cosmology theories, see \cite{elizaldev} and references therein. So, at
least for $\mu =0,$ by introducing a conformal factor $\omega $ before $%
h_{3}, h_{4}$ in above formulas, re--defining the generating
functions, and for small off--diagonal coefficients, we reproduce
nonholonomic deformations of $\Lambda $CDM universes.

The metrics (\ref{excosms}) do not have, in general, a simple physical
interpretation. Choosing the integration constants, we can extract (for
instance) Kasner type solutions with dynamical chaos etc, see examples in
\cite{afdm} and references therein. A rigorous study of
nonperturbative and nonlinear effects of such generic off--diagonal
dynamical systems even for small $\mu $ is necessary (this is a matter for our further
research). Here we note that the nonholonomic nonlinear coupling with
re--definition of generating functions by formulas (\ref{genfunct}), and by
off--diagonal coefficients of (\ref{excosms}), encodes  geometric and
physical data for MGFT into effective Einstein spaces. This follows from the
fact that such solutions are equivalent (up to frame/coordinate transforms)
to the equations $\mathbf{\check{R}}_{\mu \nu }=\check{\Lambda}\mathbf{%
\check{g}}_{\alpha \beta }.$ This motivates equivalent re--definitions of
sources $\mathbf{\Upsilon }_{\mu \nu }\rightarrow \mathbf{\hat{\Upsilon}}%
_{\mu \nu }\rightarrow (~^{m}\check{\Lambda}+~^{f}\check{\Lambda}+~^{\mu }%
\check{\Lambda}\ )\mathbf{\check{g}}_{\alpha \beta }$ as we supposed in (\ref%
{dsourc}). Considering solitonic configurations, we can polarize or "open"
for a period of time some modes of massive gravity and then "switch off"
such interactions and "pump" certain induced $f $--modified effects into
off--diagonal coefficients of Einstein metrics with redefined cosmological
constants and generating functions.

\vskip5pt

\textit{Scale factors and off--diagonal deformations of FLRW metrics: }Let
us introduce a new time coordinate $\widehat{t}$, where $t=t(x^{i},\widehat{t%
})$ and $\sqrt{|h_{4}|}\partial t/\partial \widehat{t}$, and a scale factor $%
\widehat{a}(x^{i},\widehat{t})$ when the d--metric (\ref{excosms}) can be
represented in the form%
\begin{equation}
ds^{2}=\widehat{a}^{2}(x^{i},\widehat{t})[\eta _{i}(x^{k},\widehat{t}%
)(dx^{i})^{2}+\widehat{h}_{3}(x^{k},\widehat{t})(\mathbf{e}^{3})^{2}-(%
\widehat{\mathbf{e}}^{4})^{2}],  \label{scaledm}
\end{equation}%
where $\eta _{i}=\widehat{a}^{-2}e^{\psi },\widehat{a}^{2}\widehat{h}%
_{3}=h_{3},\mathbf{e}^{3}=dy^{3}+\partial _{k}n~dx^{k},\widehat{\mathbf{e}}%
^{4}=d\widehat{t}+\sqrt{|h_{4}|}(\partial _{i}t+w_{i}).$ Small off--diagonal
deformations can be modelled with a small parameter $\varepsilon ,$ with $%
0\leq \varepsilon <1,$ where
\begin{equation}
\eta _{i}\simeq 1+\varepsilon \chi _{i}(x^{k},\widehat{t}),\partial
_{k}n\simeq \varepsilon \widehat{n}_{i}(x^{k}), \sqrt{|h_{4}|}(\partial
_{i}t+w_{i})\simeq \varepsilon \widehat{w}_{i}(x^{k},\widehat{t}).
\label{smalld}
\end{equation}

We can choose a subclass of generating functions and sources when $\widehat{a%
}(x^{i},\widehat{t})\rightarrow $ $\widehat{a}(t),\widehat{h}_{3}(x^{i},%
\widehat{t})\rightarrow \widehat{h}_{3}(\widehat{t})$ etc. Such conditions,
or of type (\ref{smalld}), have to be imposed after a locally anisotropic
solution was constructed in explicit form. This results in new classes of
solutions even in diagonal limits because of generic nonlinear and
nonholonomic character of off--diagonal systems in MGFT. For $\varepsilon
\rightarrow 0$ and $\widehat{a}(x^{i},\widehat{t})\rightarrow $ $\widehat{a}%
(t),$ we obtain scaling factors which are very different from those in
Friedmann--Lema\^ itre--Roberstson--Worker, FLRW, cosmology with GR
solutions. Nevertheless, they mimic such cosmological models with
re--defined interaction parameters and possible small off--diagonal
deformations of cosmological evolution for modified gravity theories as we
analyzed in details in \cite{elizaldev}. In this work, we consider effective
sources encoding contributions from massive gravity, with $\widehat{a}^{2}%
\widehat{h}_{3}=\frac{\check{\Phi}^{2}}{4|\check{\Lambda}|},$ where $\frac{%
\check{\Phi}^{2}}{\Phi ^{2}}=\frac{|~^{m}\Upsilon +\ ^{f}\Upsilon +\ ^{\mu
}\Upsilon |+\int dt\ \Phi ^{2}|~^{m}\Upsilon +\ ^{f}\Upsilon +\ ^{\mu
}\Upsilon |^{\diamond }}{~^{m}\check{\Lambda}+~^{f}\check{\Lambda}+~^{\mu }%
\check{\Lambda}}$.

The generating functions, sources and parameters in these formulas determine
integral varieties (i.e. general solutions) of certain systems of nonlinear
PDE. Such values have to be fixed in  forms which results in certain
physical values compatible with experimental data. Following the procedure
from section 5 of \cite{elizaldev}, we can derive a corresponding effective
field theory, see also references therein.

\vskip5pt

\textit{Reconstructing off--diagonal cosmological models in MGFT: } Let us
consider a model when the gravitational Lagrange density (\ref{actmgft}) is
chosen $\mathbf{f}(\mathbf{\tilde{R}})=\ \widehat{\mathbf{R}}+\mathbf{M}%
(~^{\mu }\mathbf{T),}$ where $~^{\mu }\mathbf{T:=T+}2~\mu ^{2}(3-tr\sqrt{%
\mathbf{g}^{-1}\mathbf{q}}-\det \sqrt{\mathbf{g}^{-1}\mathbf{q}}).$ We
denote $~^{1}\mathbf{M:=dM/d}~^{\mu }\mathbf{T}$ and $\widehat{H}:=\widehat{a%
}^{\diamond }/\widehat{a}$ for a limit $\widehat{a}(x^{i},\widehat{t}%
)\rightarrow \widehat{a}(t)$ taken for a solution (\ref{scaledm}) and
consider that an observer is in a nonholonomic basis (\ref{nadif}) with $%
N_{i}^{a}=\{n_{i},w_{i}(t)\}$ for a nontrivial off-diagonal vacuum with
effective polarizations $\eta _{\alpha }(t).$ It should be emphasized that $%
\widehat{a}(t)$ is different from $\mathring{a}(t)$ for a standard FLRW
cosmology.

The cosmological scenarios are tested in terms of the redshift $1+z=\widehat{%
a}^{-1}(t)$ for and $~^{\mu }T=~^{\mu }T(z),$ with a new \textquotedblleft
shift\textquotedblright\ derivative where (for instance, for a function $s(t)
$) $s^{\diamond }=-(1+z)H\partial _{z}.$ We can derive MGFT off--diagonal
deformed FLRW equations following the procedure considered for the formulas
(63) and (64) in \cite{elizaldev}. It is described by a set of three
equations {\small
\begin{eqnarray}
3\widehat{H}^{2}+\frac{1}{2}[\mathbf{f}(z)+\mathbf{M}(z)]-\kappa ^{2}\rho
(z) &=&0,  \notag \\
-3\widehat{H}^{2}+(1+z)\widehat{H}(\partial _{z}\widehat{H}) -\frac{1}{2}\{%
\mathbf{f}(z)+\mathbf{M}(z)+3(1+z)\widehat{H}^{2} &=&0,  \label{ceq1} \\
\rho (z)\ \partial _{z}\ \mathbf{f}&=&0.  \notag
\end{eqnarray}%
}
Re--defining the generating function, we fix the condition $\partial
_{z}\ ^{1}\mathbf{M}(z)=0$ and  satisfy the condition $\partial _{z}\
\mathbf{f}=0$ which allows nonzero densities in certain adapted frames of
references. The functional $\mathbf{M}(~^{\mu }\mathbf{T)}$ encodes degrees
of freedom of mass gravity for the evolution of the energy-density where $%
\rho =\rho _{0}a^{-3(1+\varpi )}=\rho _{0}(1+z)a^{3(1+\varpi )}.$ This is
taken for the dust matter approximation $\varpi $ and $\rho \sim (1+z)^{3}.$

Using (\ref{ceq1}), it is possible to elaborate reconstruction procedures for
nontrivial $\mu $ in a form similar to that in \cite%
{elizaldev,odintsplb,voffdmgt}. For instance, it is well known that any FLRW
cosmology can be realized in a specific $f(R)$ gravity. Here we analyze how
specific MGFTs and the FLRW cosmology can be encoded into off--diagonal
deformations. Let us introduce the \textquotedblleft
e-folding\textquotedblright\ variable $\zeta :=\ln a/a_{0}=-\ln (1+z)$  considered instead of the cosmological time $t.$ We take $\mathbf{f}(%
\mathbf{\tilde{R}})$ as in (\ref{actmgft}), use $\ \widehat{\mathbf{\Upsilon
}}(x^{i},\zeta )=~^{m}\Upsilon (x^{i},\zeta )+\ ^{f}\Upsilon (x^{i},\zeta
)+\ ^{\mu }\Upsilon (x^{i},\zeta )$ instead of (\ref{dsourc}) and
parameterize the geometric objects with dependencies on $(x^{i},\zeta )$ (in
particular, only on $\zeta $), for corresponding generating functions (\ref%
{genfunct}), where $\partial _{\zeta }=\partial /\partial \zeta $ with $%
s^{\diamond }=\widehat{H}\partial _{\zeta }s$ for any function $s.$ The
matter energy density $\rho $ is  (\ref{ceq1}).

With respect to N-adapted frames(\ref{nadif}), we can repeat all
computations leading to Eqs.~(2)-(7) in \cite{odintsplb} and prove that a MGFTs with $\mathbf{f}(\mathbf{\tilde{R}})$  realize a
FLRW like cosmological model. The nonholonomic field equation corresponding
to the first FLRW equation is
\begin{equation*}
\mathbf{f}(\mathbf{\tilde{R}})=(\widehat{H}^{2}+\widehat{H}\ \partial
_{\zeta }\widehat{H})\partial _{\zeta }[\mathbf{f}(\mathbf{\tilde{R}})]- 36
\widehat{H}^{2}\left[ 4\widehat{H}+(\partial _{\zeta }\widehat{H})^{2}+%
\widehat{H}\partial _{\zeta \zeta }^{2}\widehat{H}\right] \partial _{\zeta
\zeta }^{2}\mathbf{f}(\mathbf{\tilde{R}})\mathbf{]+}\kappa ^{2}\rho .
\end{equation*}
Introducing an effective quadratic Hubble rate, $\tilde{\kappa} (\zeta ):=%
\widehat{H}^{2}(\zeta ),$ where $\zeta =\zeta (\mathbf{\tilde{R}})$ for
certain parameterizations, this equation transforms into
\begin{equation}
\mathbf{f} =-18\tilde{\kappa} (\zeta )[\partial _{\zeta \zeta }^{2}\tilde{%
\kappa} (\zeta )+4\partial _{\zeta }\tilde{\kappa} (\zeta )]\frac{d^{2}%
\mathbf{f}}{d\mathbf{\tilde{R}}^{2}}+ 6\left[ \tilde{\kappa} (\zeta )+\frac{1%
}{2}\partial _{\zeta }\tilde{\kappa} (\zeta )\right] \frac{d\mathbf{f}}{d%
\mathbf{\tilde{R}}} +2\rho _{0}a_{0}^{-3(1+\varpi )}a^{-3(1+\varpi )\zeta (%
\widehat{\mathbf{R}})}.  \label{flem}
\end{equation}%
Off-diagonal cosmological models are determined by metrics of type (\ref%
{scaledm}),  $t\rightarrow \zeta ,$ and a functional $\mathbf{f}(%
\mathbf{\tilde{R}})$  used for computing $\widehat{\mathbf{\Upsilon }}$
and $\check{\Phi}.$ Such nonlinear systems can be described effectively by
the field equations for an (nonholonomic) Einstein space $\mathbf{\check{R}}%
_{\ \beta }^{\alpha }=\check{\Lambda}\delta _{\ \beta }^{\alpha }.$ The
value $d\mathbf{f/}d\mathbf{\tilde{R}}$ and higher derivatives vanish for
any functional dependence $\mathbf{f}(\check{\Lambda})$ with $\partial
_{\zeta }\check{\Lambda}=0.$ Even we work with off--diagonal configurations,
the recovering procedure simplifies substantially in such cases.

\vskip5pt

\textit{An example of reconstruction of MGFT and nonholonomically deformed
Einstein spaces reproducing the }$\Lambda $\textit{CDM era: }We consider any
$\widehat{a}(\zeta )$ and $\widehat{H}(\zeta )$ determined by an
off-diagonal solution (\ref{scaledm}), with respect to correspondingly
N-adapted frames. The analog of FLRW equation for ${\Lambda }$CDM cosmology
is
\begin{equation}
3\kappa ^{-2}\widehat{H}^{2}=3\kappa ^{-2}H_{0}^{2}+\rho _{0}\widehat{a}%
^{-3}=3\kappa ^{-2}H_{0}^{2}+\rho _{0}a_{0}^{-3}e^{-3\zeta },  \label{aux11}
\end{equation}%
where $H_{0}$ and $\rho _{0}$ are fixed to be certain constant values. Such
assumptions are considered after the coefficients of off-diagonal solutions
are found and where the dependencies on $(x^{i},\zeta )$ are changed into
dependencies on $\zeta .$ The values with "hat" are generated via a
corresponding re-definition of the generating functions and the effective
sources. The first term on the rhs is related to an effective cosmological
constant $\check{\Lambda}$ (\ref{dsours}) which appears in  re-definition (%
\ref{genfunct}). For this model, the second term in (\ref{aux11}) describes,
in general, an inhomogeneous distribution of cold dark mater (CDM). The
similarity with the diagonalizable cosmological models in GR is kept if $\check{\Lambda}=12H_{0}^{2}$ to survive in the limit $%
w_{i},n_{i}\rightarrow 0,$ for certain approximations of type (\ref{smalld}).

The effective quadratic Hubble rate and the modified scalar curvature, $%
\mathbf{\tilde{R}}$, are computed using (\ref{aux11}), respectively,
\begin{equation*}
\tilde{\kappa} (\zeta ):= H_{0}^{2}+\kappa ^{2}\rho _{0}a_{0}^{-3}e^{-3\zeta
}\mbox{ and } \mathbf{\tilde{R}}=3\partial _{\zeta }\tilde{\kappa} (\zeta
)+12\tilde{\kappa} (\zeta )=12H_{0}^{2}+\kappa ^{2}\rho
_{0}a_{0}^{-3}e^{-3\zeta }.
\end{equation*}%
(\ref{flem}) transforms into%
\begin{equation}
X(1-X)\frac{d^{2}\mathbf{f}}{dX^{2}}+[\chi _{3}-(\chi _{1}+\chi _{2}+1)X]%
\frac{d\mathbf{f}}{dX}-\chi _{1}\chi _{2}\mathbf{f}=0,  \label{gauss}
\end{equation}%
for certain constants, for which $\chi _{1}+\chi _{2}=\chi _{1}\chi
_{2}=-1/6 $ and $\chi _{3}=-1/2$ where $3\zeta =-\ln [\kappa ^{-2}\rho
_{0}^{-1}a_{0}^{3}(\mathbf{\tilde{R}}-12H_{0}^{2})]$ and $X:=-3+\mathbf{%
\tilde{R}}/3H_{0}^{2}.$ The solutions of such equations with constant
coefficients and for different types of scalar curvatures were found in \cite%
{odintsplb} and \cite{elizaldev} as Gauss hypergeometric functions.
Similarly, we denote $\mathbf{f}=F(X):=F(\chi _{1},\chi _{2},\chi _{3};X),$
where for some constants $A$ and $B$, \
\begin{equation*}
F(X)=AF(\chi _{1},\chi _{2},\chi _{3};X)+ BX^{1-\chi _{3}}F(\chi _{1}-\chi
_{3}+1,\chi _{2}-\chi _{3}+1,2-\chi _{3};X).
\end{equation*}%
This provides a proof of the statement that MGFT can indeed describe ${%
\Lambda }$CDM scenarios without the need of an effective cosmological
constant.

\vskip5pt

\textit{Final remarks:}\ One of the most interesting results of applications
of the AFDM \cite{afdm} to nonlinear MGFTs systems is that via
re--definition of generating functions and effective sources we can mimic $f$%
--modifications and massive gravity effects. This is possible by modelling
modified theories via off--diagonal interactions in effective Einstein
spaces. Such models are generically nonlinear, parametric and with respect
to nonholonomic frames which allows to decouple and integrate the associated
PDEs in general forms.

There is a proof of absence of FLRW cosmology in massive gravity (see section 2.1 in \cite%
{amico}). It proof follows for homogeneous and isotropic
ansatz for metrics in certain models of massive theory. In this paper, we studied more
general constructions both for modified gravity functionals and
off--diagonal locally anisotropic and inhomogeneous metrics. Our solutions
describe massive gravity effects encoded both in effective matter sources
and in off--diagonal deformations. Even for very special cases when $f(%
\tilde{\mathbf{R}})$ is linear on $\tilde{\mathbf{R}}$ such
contributions are not trivial because such a scalar curvature is computed  not for
the Levi--Civita connection but for a nonholonomically deformed ansatz.
Considering holonomic configurations, we can reproduce the general results
\cite{sarid} or model cosmological scenarios from \cite{odintsr}. For
nonlinear systems, it is very important  when certain assumptions
and additional constraints are considered. If some "simplifications" or
approximations are made at the very beginning, we formulate certain
conclusions about properties of a theory and even follow a procedure of
finding of solutions. But  we can also eliminate a number of other
types of solutions and various nonlinear characteristics. In our approach,
we elaborated a more general and more realistic model with generic
off--diagonal effects with certain stability configurations and
off--diagonal modifications of FLRW cosmology generated by effective sources
in nonlinear massive gravity.

\vskip5pt

\textbf{Acknowledgments:\ } The work is partially supported by the Program
IDEI, PN-II-ID-PCE-2011-3-0256 and a visiting research program at CERN. SV
is grateful to N. Mavromatos, P. Stavrinos and S. Rajpoot for important discussions and substantial support.



\begin{thebibliography}{99}
\bibitem{sarid} Yi-Fu Cai, F. Duplessis and E. N. Saridakis, arXiv: 1307.7150

\bibitem{odintsr} J. Kluso\v{n}, S. Nojiri and S. D. Odintsov, Phys. Lett.
\textbf{B 726} (2013) 918; S. Nojiri and S. D. Odintov, Phys. Lett. \textbf{%
B 716} (2012) 377; S. Nojiri, S. D. Odintsov and N, Schirai, JCAP\textbf{\
1305} (2013) 020

\bibitem{drgt} C. de Rham and G. Gabadadze, Phys. Rev. \textbf{D 82} (2010)
044020; C. de Rham, G.\ Gabadadze and A. J. Tolley, Phys.\ Rev. Lett.
\textbf{106} (2011) 231101

\bibitem{ghost} D. G. Boulware and S. Deser, Phys.\ Rev. \textbf{D 6} (1972)
3368; J. Kluson, Phys. Rev. \textbf{D 86} (2012) 044024; S.\ F. Hassan and
R. A. Rosen, JHEP \textbf{1204} (2012) 123; A. Golovnev, Phys. Lett.\
\textbf{B 707} (2012) 4004

\bibitem{mgt} S. Nojiri and S. D. Odintsov, eConf C (2006) 0602061 [Int. J.
Geom. Meth. Mod. Phys. \textbf{4} (2007) 115]; Phys.\ Rept. \textbf{505}
(2011) 59; S. Capozziello and V. Faraoni, Beyond Einstein gravity: A survey
of gravitational theories for cosmology and astrophysics (Springer, 2010)

\bibitem{stelle} K. S. Stelle, Phys. Rev. \textbf{16} (1977) 953; S. Deser,
J. H. Kay and K. S. Stelle, Phys. Rev. Lett. \textbf{38} (1977) 527

\bibitem{mgt1} E. N. Saridakis, Class. Quant. Grav. \textbf{30} (2013)
075003; Y. -I. Zhang, R. Saito and M. Sasaki, JCAP \textbf{1302} (2013) 029;
M. Mohseni, JCAP \textbf{1211} (2012) 023; K. Hinterbichler, J. Stokes and
M. Trodden, Phys. Lett. \textbf{B 725} (2013) 1; M. Andrews, G. Goon, K.
Hinterbichler, J. Stokes and M. Trodden, Phys. Rev. Lett. \textbf{111}
(2013) 061107; R. Gannouji, M. W. Hossain, M. Sami and E. N. Saridakis,
Phys.\ Rev. \textbf{D 87} (2013) 123536; S. Capozziello and P. Martin-Morun,
Phys. Lett. \textbf{B 719} (2013) 14; J. Kluson, Phys. Rev. \textbf{D86}
(2012) 044024

\bibitem{bmcm} T. Damour, I. I. Kogan and A. Papazoglou, Phys. Rev. D 66
(2002) 104025; M. S. Vokov, JHEP 1201 (2012) 035; M. von Strauss, A.
Schmidt-May, J. Enangder, E. Mortsell and S. F. Hassan, JCAP 1203 (2012)
042; M. Berg, I. Buchberger, J. Enander, E. Mortsell and S. Sjors, JCAP 1212
(2012) 021

\bibitem{afdm} S. Vacaru, Eur. Phys. J. \textbf{C 73} (2013) 2287; Europhys.
Letters \textbf{96} (2011) 5001; IJGMMP \textbf{8} (2011) 9; J. Phys.: Conf.
Ser. \textbf{543} (2013) 012021; J. Math. Phys. \textbf{46} (2005) 042503
JHEP \textbf{04} (2001) 009; S. Vacaru and D. Singleton, Class. Quant. Grav.
\textbf{19} (2002) 2793

\bibitem{vmgbmconf} S. Vacaru, IJGMMP \textbf{\ 11} (2014) 1450032

\bibitem{elizaldev} E. Elizalde and S. Vacaru, arXiv: 1310.6868

\bibitem{odintsplb} S. Nojiri, S. D. Odintsov and D. Saez-Gomez, Phys. Lett.
\textbf{B681} (2009) 74

\bibitem{voffdmgt} P. Stavrinos and S. Vacaru, Class. Quant. Grav. \textbf{%
30 }(2013) 055012;\ S. Vacaru, arXiv: 1305.1876

\bibitem{amico} G. D'Amico, C. de Rham, S. Dubovsky, G.\ Gabadadze, D.\
Pirtskhalava and A. J.\ Tolley, Phys. Rev. \textbf{D 84} (2011) 124046
\end{thebibliography}
\end{document}